# Quantization of Paths in Space-Time and Non-Perturbative Quantum Gravity

James Moffat [1] and Charles Wang

*Dept of Physics, University of Aberdeen*

## Abstract

Building on previous work on the quantization of paths in space - time, we have developed a mathematically coherent theory addressing a number of open questions concerning Loop Quantum Gravity. Our approach develops a discrete space-time and shows that macroscopic space-time is a renormalization limiting form. Weaving together a number of our previous results we then prove that quantum states invariant under either an external group of local diffeomorphisms of space-time or, by contrast, quantum states invariant under the internal action of a compact Lie group are 'common', in a well-defined sense. These form the building blocks of invariant fields and Lagrangians. A form of $N$ = 1 Supersymmetry and non-commutative space – time naturally emerges, which predicts a spin- 2 massless graviton and its companion gravitino.

## Spinors and Spin Networks

A left-handed Weyl 2-spinor is an element of a 2-dimensional vector space $F$ with a basis of Clifford variables denoted $\psi_A$ (A = 1,2).

Given such a 2-spinor $\psi_A$, the 2x2 matrix M acting on $\psi_A$ gives rise to another 2-spinor $\psi'_A = M_A^{\,B}\psi_B$, summing over repeated indices.

Left-handed and right-handed Weyl spinors are related as follows;

$$\psi_A \in F \text{ implies } \psi_A^* = \bar{\psi}_{\dot{A}} \in \dot{F}$$

[1] Corresponding author, email; jamesmoffat48@gmail.com

The representations $M \to M$ and $M \to \left(M^{-1}\right)^T$ have the same dimensionality and are therefore unitarily equivalent; there is a unitary matrix $\varepsilon = \varepsilon_{AB}$ such that for all M,

$\varepsilon^{=1} M \varepsilon = \left(M^{-1}\right)^T$ where $\varepsilon = \begin{pmatrix} 0 & -1 \\ 1 & 0 \end{pmatrix}$.

Defining $\varepsilon^{AB}$ by $\left(\varepsilon^{AB}\right)^{-1} = \begin{pmatrix} 0 & -1 \\ 1 & 0 \end{pmatrix}$ we have $\varepsilon^{AB} = \begin{pmatrix} 0 & 1 \\ -1 & 0 \end{pmatrix}$.

The matrices $\varepsilon_{AB}$ and $\varepsilon^{AB}$ are tensor objects which can be used to raise and lower indices in the usual vector and tensor calculus, within which left-handed 'chiral' Weyl 2-spinors are covariant vectors, and right-handed chiral Weyl 2-spinors are contravariant vectors. Denoting the dual space as *F**. the following calculation illustrates this 'spinor calculus' for $\psi \in F^*$ and $\chi \in F$ ;

$$\psi\chi = \psi^A \chi_A = \varepsilon^{AB}\psi_B\chi_A = \varepsilon^{12}\psi_2\chi_1 + \varepsilon^{21}\psi_1\chi_2 = \psi_2\chi_1 - \psi_1\chi_2$$

We also define the matrix $\bar{\varepsilon}$ such that the representations

$M \to M^*$ and $M \to \left(M^{*-1}\right)^T$ are equivalent, so that $\bar{\varepsilon}^{-1} M^* \bar{\varepsilon} = \left(M^{*-1}\right)^T$ for all $M^*$.

We now consider in this Weyl 2 - spinor context the Dirac 4x4 spin matrix representation;

$\gamma^\mu = \begin{pmatrix} 0 & \sigma^\mu \\ \bar{\sigma}^\mu & 0 \end{pmatrix}$ where $\sigma^\mu$ ($\mu$=0,1,2,3) are the Pauli 2 x 2 spin matrices.

In this representation, we can show directly that $i\gamma^0\gamma^1\gamma^2\gamma^3 = \begin{pmatrix} -I_{2x2} & 0 \\ 0 & I_{2x2} \end{pmatrix}$ where $I_{2x2}$ is the 2x2 identity matrix $\begin{pmatrix} 1 & 0 \\ 0 & 1 \end{pmatrix}$.

This Dirac spin operator representation acts on a 4-dimensional vector space. Each element of this vector space suitably normalized, is a quantum ket state $|\psi>$; a pure state. Defining $\gamma^5$ to be the composite matrix operator $i\gamma^0\gamma^1\gamma^2\gamma^3$, we can write this state $|\psi>$ in the form $|\psi> = \frac{1}{2}(I_{4x4} - \gamma^5)|\psi> + \frac{1}{2}(I_{4x4} + \gamma^5)|\psi>$. These correspond to the left handed and right handed 2-spinor components of the state;

$\psi_L = \frac{1}{2}(I_{4x4} - \gamma^5)\psi$ and $\psi_R = \frac{1}{2}(I_{4x4} + \gamma^5)\psi$

So that $\psi = \psi_L + \psi_R$.

Now we know that $(I_{4x4} - \gamma^5) = \begin{pmatrix} I_{2x2} & 0 \\ 0 & I_{2x2} \end{pmatrix} - \begin{pmatrix} -I_{2x2} & 0 \\ 0 & I_{2x2} \end{pmatrix} = \begin{pmatrix} 2I_{2x2} & 0 \\ 0 & 0 \end{pmatrix}$

Similarly, $I_{4x4} + \gamma^5 = \begin{pmatrix} 0 & 0 \\ 0 & 2I_{2x2} \end{pmatrix}$.

Thus, the four vector $\psi_L$ has only two non-zero elements (the first two) and is of the form $\psi_L = \begin{pmatrix} \psi_1 \\ \psi_2 \\ 0 \\ 0 \end{pmatrix}$. Similarly, the four vector $\psi_R$ has only two non-zero elements (the last two) and is defined as $\psi_R = \begin{pmatrix} 0 \\ 0 \\ \bar{\chi}^{\dot{1}} \\ \bar{\chi}^{\dot{2}} \end{pmatrix}$. These then correspond to the left handed chiral Weyl 2-spinor $\begin{pmatrix} \psi_1 \\ \psi_2 \end{pmatrix}$ and the right handed chiral Weyl 2-spinor $\begin{pmatrix} \bar{\chi}^{\dot{1}} \\ \bar{\chi}^{\dot{2}} \end{pmatrix}$. In four dimensions we have $\psi = \begin{pmatrix} \psi_L \\ \psi_R \end{pmatrix}$ which is an element of the vector space $E = F \oplus \dot{F}^*$.

Assuming planar isotopy, it is possible to associate various locally deformable lines in the plane with Weyl 2-spinor calculations, giving rise to topological structures called spin networks. As a simple example the tensors $\bar{\varepsilon} = \varepsilon^{\dot{A}\dot{B}} = \begin{pmatrix} 0 & 1 \\ -1 & 0 \end{pmatrix}$ and $\bar{\varepsilon}^{-1} = \varepsilon_{\dot{A}\dot{B}} = \begin{pmatrix} 0 & -1 \\ 1 & 0 \end{pmatrix}$ correspond to;

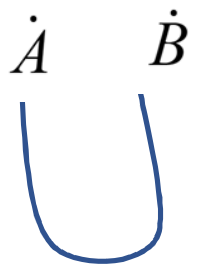

and

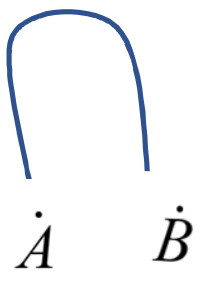

Their product is then a closed loop which corresponds to a scalar λ times the identity matrix with λ =1 in this case. The advantage of this approach is that spinor calculations become a sequence of potentially simpler topological transformations with connections to knot theory.

**Computational Spin Networks**

Building on these ideas, we define a *computational spin network* to be a finite quiver consisting of a directed graph with *n* nodes, where the nodes represent entangled spin inputs and a directed link between two nodes corresponds to a quantum gate, as we now discuss.

If we represent the basis spin ±½ eigenvectors as the column vectors $0 >= \begin{pmatrix} 1 \\ 0 \end{pmatrix}$ and $1 >= \begin{pmatrix} 0 \\ 1 \end{pmatrix}$, then the NOT quantum gate which switches $|0>$ to |1> and |1> to |0> corresponds to the unitary matrix $U = \begin{pmatrix} 0 & 1 \\ 1 & 0 \end{pmatrix}$. It is easy to check that $U|0>=|1>$ and $U|1>=|0>$. All quantum gates correspond in this way to multiplying the input state vector by a unitary matrix.

To ease notational clutter, in all that follows we denote the joint tensor product state of the spins |x> and |y> as |xy>.

Given the entangled state, $|\beta(0,0)> = \frac{1}{\sqrt{2}}(|00> + |11>)$ and $|y_0> = \frac{1}{\sqrt{2}}\left(\alpha|0> + \beta|1>\right)$;

Let $|y_1> = |y_0> \otimes |\beta(0,0)> = \frac{1}{2}\alpha|0> \otimes |\beta(0,0)> + \beta|1> \otimes |\beta(0,0)>$

$= \frac{1}{2}\left(\alpha|0> \otimes (|00> + |11>) + (\beta|1> \otimes |00> + |11>\right)$

$\equiv \frac{1}{2}\left(\alpha|000> + \alpha|011>) + (\beta|100> + \beta|111>\right)$

Define $|y_2> = U(CNOT)|y_1>$ mapping;

$|00\delta> \to |00\delta>$; $|01\delta> \to |01\delta>$; $|10\delta> \to |11\delta>$ and $|11\delta> \to |10\delta>$ where $\delta \in \{0,1\}$

$\Rightarrow |y_2> = \frac{1}{2}\left(\alpha|000> + \alpha|011>) + (\beta|110> + \beta|101>\right)$

Replacing $|y_0> = \frac{1}{\sqrt{2}}\left(\alpha|0> + \beta|1>\right)$ by using the Hadamard unitary gate U(H) we have

$U(H)|y_0> = \frac{1}{\sqrt{2}}\alpha U(H)|0> + \beta U(H)|1> = \frac{1}{2}\{(\alpha|0> + \alpha|1>) + (\beta|0> - \beta|1>)\}$

Then we have;

$$|y_2> = \frac{1}{2}\left(\alpha|000> + \alpha|011>) + (\beta|110> + \beta|101>\right)$$

$$= \frac{1}{2}\left((\alpha|0>(|00> + |11>) + \beta|1>\left(|10> + |01>\right)\right)$$

$$\Rightarrow |y_3> = \frac{1}{2}\left\{\alpha U(H)|0>(|00> + |11>) + \beta U(H)|1>\left(|10> + |01>\right)\right\}$$

$$= \frac{1}{4}\left\{\left(\alpha|0> + \alpha|1>\right)(|00> + |11>) + (\beta|0> - \beta|1>)\left(|10> + |01>\right)\right\}$$

$$= \frac{1}{4}\{\alpha|000> + \alpha|011> + \alpha|100> + \alpha|111> + \beta|010> + \beta|001> - \beta|110> - \beta|101>\}$$

$$= \frac{1}{4}\{|00>(\alpha|0> + \beta|1>) + |01>(\alpha|1> + \beta|0> + |10>(\alpha|0> - \beta|1> + |11>(\alpha|1> - \beta|0>\}$$

This can then be interpreted in terms of the original vector $|y_0>$ as;

$$\frac{1}{4}\{|00>U(00)|y_0>+|01>U(01)|y_0>+|10>U(10)|y_0>+|11>U(11)|y_0>\}$$

Comparing the two expressions implies that;

$$U(00)=I_{2x2}=\begin{pmatrix}1&0\\0&1\end{pmatrix};U(01)=\begin{pmatrix}0&1\\1&0\end{pmatrix}^{-1};U(10)=\begin{pmatrix}1&0\\0&-1\end{pmatrix}^{-1};U(11)=\begin{pmatrix}0&1\\-1&0\end{pmatrix}^{-1}$$

These are all real unitary matrices thus their inverse is simply the transpose matrix in each case;

$$U(00)=I_{2x2}=\begin{pmatrix}1&0\\0&1\end{pmatrix};U(01)=\begin{pmatrix}0&1\\1&0\end{pmatrix};U(10)=\begin{pmatrix}1&0\\0&-1\end{pmatrix};U(11)=\begin{pmatrix}0&-1\\1&0\end{pmatrix}$$

Thus, we can transport a Quantum State along such a computational path.

Note that $U(00)=\sigma^0 \quad U(01)=\sigma^1 \quad U(10)=\sigma^3 \quad iU(11)=\sigma^2$ where $\left\{\sigma^j; j=0,1,2,3\right\}$ are the Pauli spin matrices.

We identify this transport from node to node of the computational spin network with the action of a sequence of discrete translations in space-time from one node to the next neighboring node. A discrete path in space-time can be then be considered as a series of applications of the translation subgroup of the Poincare group.

**Discrete Paths in Space-Time**

We start by considering classical phase space. Given a dynamical system, entropy is defined through considering the phase space of the system. The emergent behaviour of this classical system gives rise to regions of phase space, each corresponding to similar macro-level behaviour. The entropy of such a coarse-grained region is a measure of all the different micro-configurations constituting that region. A system starting in a low entropy state will tend to wander into larger coarse-grained volumes; hence thermodynamic entropy tends to increase over time if the system is isolated, giving rise to the second law of thermodynamics. The structure of classical phase space is such that each set of initial conditions $(x_\mu, p_\mu)$ generates a unique solution $S(x_\mu, p_\mu)$. For a Hamiltonian system it is possible to reformulate classical mechanics as a symplectic vector space of solutions, or an equivalent set of initial conditions of location and momentum, equipped with a bilinear form Ω which ultimately derives from Hamilton's equations of motion;

With $\xi \equiv (x, p)$ a 6-dimensional vector we have $\dfrac{\partial \xi}{\partial t} = \Omega \dfrac{\partial H}{\partial \xi}$ where $\Omega = \begin{bmatrix} 0_{3\times3} & I_{3\times3} \\ -I_{3\times3} & 0_{3\times3} \end{bmatrix}$.

This is of the form of a symplectic vector space $V \oplus V^*$ where *V* is a real finite vector space with dual *V**. The skew-symmetric rank 2 tensor Ω then takes the general form;

$$\Omega(\ x \oplus \eta\ ,\ x' \oplus \eta'\ )\ =\ \eta' \bullet x\ -\ \eta \bullet x'\ .$$

In our case *V* is the configuration space, *V** the (dual) momentum space and $V \oplus V^*$ the phase space, a product vector bundle over *V* with fibre *V**. By choosing values such as (1,0,0,0,0,0) we can pull out elements;

$$\Omega(\ x \oplus \eta\ ,\ 0 \oplus \eta'\ )\ =\ \eta' \bullet x\ = \eta'_1\ .$$

The Dirac canonical quantisation of elements of phase space such as $\eta'_1$ is equivalent to the canonical quantisation; $\Omega \rightarrow \hat{\Omega}$ as a (not necessarily bounded) linear operator, and this form of canonical quantisation extends smoothly to countably infinite phase space [1].

Given the canonical quantisation; $\Omega \rightarrow \hat{\Omega}$ we can form the Weyl unitaries $\hat{W} = \exp i\hat{\Omega}$ . Then closure of linear combinations of these unitaries and their adjoints in the normed operator topology is the Weyl algebra.

The extension of Ω to the space of solutions allows us to define an inner product on *S* as $\langle S(1), S(2) \rangle = -i\Omega\left(S(1)^*, S(2)\right)$ where $S(1)^*$ is the complex conjugate solution. It turns out [1] that this defines an inner product on *S* relative to which a one particle Hilbert space can be defined. For a quantum system of bosonic harmonic oscillators, we can then assemble a symmetric tensor product Fock space in the usual way, using creation and annihilation operators.

An example of the Weyl form in a two dimensional locally flat space-time is now given for a local algebra *O(D)*, having a representation as observables acting on the Hilbert space $L^2(x,t)$ with Lebesgue measure.

*For a small increment of space-time* $(\delta x, \delta t)$ *we consider the Poincare Translation subgroup element* $T(\delta x, \delta t):(x,t) \to (x+\delta x, t+\delta t)$ *and define;*

$$U_{T(\delta x,\delta t)} f(x,t) = f(x-\delta x, t-\delta t) \text{ for } f(x,t) \in L^2(x,t)$$

*Then U is a local group homomorphism of the translation group T as observables acting on* $L^2(x,t)$. *Define also;*

$$\begin{aligned} &V_{\delta t} = U_{T(0,\delta t)}; V_{\delta x} = U_{T(\delta x,0)} \\ &\Rightarrow V_{\delta t} V_{\delta t'} f(x,t) = U_{T(0,\delta t)} U_{T(0,\delta t')} f(x,t) = f(x, t-\delta t' - \delta t) \\ &= U_{T(0,\delta t'+\delta t)} f(x,t) = V_{\delta t+\delta t'} f(x,t) \Rightarrow V_{\delta t} V_{\delta t'} = V_{\delta t+\delta t'} \end{aligned}$$

*A similar result applies for* $V_{\delta x}$, *by symmetry.*

*Now introduce a deformation of the form;* $T(\delta x, 0) \to Z_{\delta x} f(x,t) = \exp(it\delta x) f(x-\delta x, t)$.

*The mappings V and Z are unitary representations on* $L^2(x,t)$ *and so also is their product* $V \times W : (\delta x, \delta t) \to T(\delta x, \delta t) \to V_{\delta t} Z_{\delta x}$

*Then we have;*

$$\begin{aligned} V_{\delta t} Z_{\delta x} f(x,t) &= V_{\delta t} \exp(it\delta x) f(x-\delta x, t) \\ &= \exp(it\delta x) f(x-\delta x, t-\delta t) \\ Z_{\delta x} V_{\delta t} f(x,t) &= Z_{\delta x} f(x, t-\delta t) \\ &= \exp(i(t-\delta t)\delta x) f(x-\delta x, t-\delta t) \\ &= \exp(-i\delta t \delta x) V_{\delta t} Z_{\delta x} f(x,t) \\ Z_{\delta x} V_{\delta t} &= \exp(-i\delta t \delta x) V_{\delta t} Z_{\delta x} \end{aligned}$$

*Then* $T(\delta x, \delta t) \to (Z_{\delta x}, V_{\delta t})$ *is a local Weyl representation of the CCR on* $L^2(x,t)$. By the Stone-von Neumann theorem, the resulting C*-algebra and its weak closure as a von Neumann algebra must be unitarily isomorphic to Wald's equivalent 'algebraic approach' to quantum field construction and his Weyl Algebra, since we can assume all relevant Hilbert spaces are separable in application to observed real systems [1].

This example indicates that a continuous local group homomorphism from a neighbourhood of the identity of T to a neighbourhood of the identity of the set of observables in *O(D*) exists as a Weyl algebra. It can be easily extended to 4 dimensions by replacing x by the 3-vector **x** = $(x_1, x_2, x_3)$.

A discrete path in space-time, as we have so far generated it, can be considered as a set of linked causally directed intervals each of fractal dimension 1 in renormalized smooth space-time. More formally, we define the path as a series of *n* linked increments **a(***j***)** with varying direction relative to a local forward light cone, such that the path begins at **x(0)** and ends at **x(1)**, with $T(\mathbf{a}(j)) : \mathbf{x} \to \mathbf{x} + \mathbf{a}(j)$ elements of the translation subgroup *T*. The total path is then generated by the product $\prod_{j=1}^{j=n} T(\mathbf{a}(j))\mathbf{x}(\mathbf{0})$ with the final end point $\mathbf{x}(\mathbf{1}) = \mathbf{x}(\mathbf{0}) + \sum_{j=1}^{j=n} \mathbf{a}(j)$. For a fixed initial point **x(0)** we can identify this path with the finite group product $\prod_{j=1}^{j=n} T(\mathbf{a}(j)) \in T$.

We can now construct a space-time, assumed non-commutative at some energy level such as the Planck regime, with non-commutative algebraic structure at each event point **x** of space-time, forming the fibre algebra **A(x)**. This structure then corresponds, we assume, to the single fibre of a principal fibre bundle. A gauge group of automorphisms corresponding to the translation subgroup T of the Poincare group acts on each fibre algebra locally, while a section through this bundle is then a quantum field of the form $\{A(x); x \in M\}$ with *M* the renormalisation limit macroscopic space-time manifold to be constructed in the next section. In addition, we assume a local algebra *O(D)* corresponding to the algebra of sections of such a principal fibre bundle with base space a finite and bounded subset of space-time, $D \subset M$. The algebraic operations of addition and multiplication are assumed defined fibrewise for this algebra of sections.

Now let there be given a continuous local group homomorphism from a neighbourhood *V* of the identity of *T* to the neighbourhood *W* of the identity of the set of observables in *O(D)* as a Weyl representation of the CCR. Then a discrete classical path *CP* in space-time can be lifted to a quantum field section *QP* through *O(D)*. This we proved in reference [2]; we also proved that the path *QP* is uniquely determined, and that there is a projection $\pi$ from the fibre bundle *O(D)* mapping the quantum field back to the path *CP*.

**Renormalisation of Discrete Paths in Space-Time**

The principle of relativity is captured within the assumptions of the Riemannian geometry of 4-manifolds, where formulae equating a tensor expression to zero remain invariant under local diffeomorphism transformations. It is a natural extension of these ideas to additionally postulate that the scales of measurement inscribed on the clocks or measuring rods used by an observer should also not be absolute. Mathematically this can be captured by the additional requirement that the tensor formulae should be invariant under transformations of scale. From this perspective a relativistic quantum system is a *scale free system.* If $\Phi$ is the function transforming system inputs to system outputs, then for a scale-free system, $\Phi$ is invariant under a change of scale.

Under the assumptions of such a scale-invariant relativity, let us consider a discrete closed loop in space-time; corresponding to two discrete non-oriented paths sharing the same end points. Then it turns out [2] that this loop is renormalisable and has a finite limit corresponding to the curve fractal dimension as a curve in macroscopic spacetime.

**Quantum States Invariant Under the Action of Local Space-Time Diffeomorphisms**

**Gravity States, the Graviton and Supersymmetric Gravitino**

We now investigate in more depth the subgroup *T* of the Poincare group consisting of translations of macroscopic space-time as a gauge group of automorphisms. We define a representation of *T* as a group of automorphisms of a local fibre algebra **A(x)** which we assume to be isomorphic to a von Neumann algebra with trivial centre acting on a separable Hilbert space, rooted at the event point x. Consider then the subgroup *T* acting on **A(x).** These actions generate the local diffeomorphisms of General Relativity. If we make the minimal assumption that this representation is weakly measurable; i.e. the mapping $g \to \langle \alpha_g(A)x, y \rangle$: is Haar-measurable for all relevant values of *A*, *x* and *y*; then the argument of [3] shows that the mapping $g \to \alpha_g$ is norm continuous. Since the translational group is both abelian and connected, it follows [4] that each $\alpha_g$ is an inner automorphism of **A** and the corresponding unitary $W_g$ has a spectrum contained in the positive half plane. In fact, we have;

$$\sigma(W_g) \subset \{z; \operatorname{Re} z \geq \beta_g\} \text{ where } \beta_g = \frac{1}{2}\left(4 - \| \alpha_g - i \|^2\right)^{\frac{1}{2}}$$

Moreover, if **S** denotes the von Neumann subalgebra generated by $\{W_g; g \in T\}$, then the set of unitaries $\{W_g; g \in T\}$ is a commuting set within **A(x).** Thus, **S** is a commutative subalgebra and contains the identity I of **A(x)**, since if *id* is the group identity then $I = W_{id}$. **A(x)** is a factor, **S** thus contains the centre Z of **A(x)**. Such a commutative quantum operator algebra is equivalent to the set of continuous functions on a compact space and this equivalence arises through the Gelfand transform;

$$A \to \hat{A} \text{ with } \hat{A}(\rho) = \rho(A) \text{ and } \rho \text{ a continous complex valued homomorphism.}$$

The carrier space $\Phi_S$ of **S** is the set of all such continuous complex valued homomorphisms on **S**. and is thus topologically a Stonean space, as is $\Phi_Z$ and the restriction map $\pi : \Phi_S \to \Phi_Z$ is a continuous surjection**.** We then have, as we will prove, a lifting

$$f : \Phi_Z \to \Phi_S \text{ with } f(\rho)|_Z = \pi \circ f(\rho) = \rho \text{ for } \rho \in \Phi_Z .$$

For each $g \in T$, we can now define the Gelfand transform of a unitary $U_g$ acting on the carrier space $\Phi_S$ of **S** as follows;

$$\hat{U}_g(\rho) = \overline{\hat{W}}_g(f(\rho|_Z))\hat{W}_g(\rho); \quad \rho \in \Phi_S$$

If we define the equivalence $\rho \approx \rho' \Leftrightarrow \rho|_Z = \rho'|_Z$ then by the extended form of the Stone-Weierstrass theorem [5] the centre Z corresponds to those elements of $\Phi_S$ constant on each equivalence class, Applying this to the Gelfand transform $\hat{V}_g(\rho) = \overline{\hat{W}}_g(f(\rho|_Z))$ it follows that $\hat{V}_g$ corresponds to an element $V_g$ of the centre. Since **A(x)** is a factor, this means that

$V_g = \nu(g)I$ with $\nu(g)$ a complex number defining a coboundary, and that for each $g \in T$; $U_g$ implements $\alpha_g$. In addition, the mapping $g \to U_g$ is a group homomorphism for if we set;

$R_{g,h} = U_g U_h U_{gh}^*$ then for an operator *A*, $R_{g,h} A R_{g,h}^* = U_g U_h U_{gh}^* A U_{gh} U_h^* U_g^* = \alpha_g \alpha_h \alpha_{gh}^{-1}(A) = A$
Hence $R_{g,h}$ is a unitary in the centre. In fact $R_{g,h} = \lambda(g,h)I$ where $\lambda(g,h)$ is a 2-cocycle.

The fact that the lifting $f : \Phi_Z \to \Phi_S$ has the property $\pi \circ f(\rho) = f(\rho)|_Z = \rho$ for $\rho \in \Phi_Z$ means that $\hat{R}_{g,h}(\rho) = \hat{R}_{g,h}(f(\rho))$ since the domain of $\hat{R}_{g,h}$ is $\Phi_Z$ . But then we have that;

$\hat{U}_g(f(\rho)) = 1 \quad \forall g \in G,$ thus $\hat{R}_{g,h}(\rho) = 1 \quad \forall \rho \in \Phi_Z$; the 2-cocycle is trivial.
Therefore $R_{g,h} = I \quad \forall g,h \in G$ and $g \to U_g$ is a group representation by unitaries in the fibre algebra, which turns out to be norm continous, due to their spectral characteristics [4].

We now show that the lifting follows from the Axiom of Choice, in the guise of the equivalent Zorn's Lemma, applied to the strange topological properties of Stonean spaces. We can set it in the context of lifting from a totally disconnected, compact Hausdorff base space B into a containing fibre bundle K having a Stonean topology.

**Lifting from the Base Space B of a Stonean Fibre Bundle K.**

We prove first the fact that the projection $\pi : K \to B$ is an open mapping if and only if for each non-trivial open subset *E* of *K*, *π(E)* is not a nowhere dense subset of the base space *B.*
One way is trivial for if *π* is an open mapping then *π(E)* is a non-trivial open set so cannot be nowhere dense.
Conversely, the Stonean topology of the fibre bundle K is compact and totally disconnected, with a basis of 'clopen' sets (i.e. sets which are both closed and open). Thus, every open set is a union of such clopen sets, and it suffices to show that if V is clopen in K, then *π(V)* is open in B. Since V is clopen, it is a closed and thus compact subset of K, and *π* is continuous, thus *π (V)* is compact. If we define $Y = \pi(V) \setminus \operatorname{int} \pi(V)$; this a closed set with empty interior thus Y is a nowhere dense set and is the image of an open set; $Y = \pi\left(V \setminus \pi^{-1}(\operatorname{int} \pi(V)\right)$, using the fact that the projection mapping π is continuous and surjective. It follows that *Y* is void and *π(V)=int π(V).*
Thus *π(V)* is open and *π* is an open mapping.

**A Unique quantum field isomorphic to the  Base Space**

Consider now the set Φ of all compact subsets S of K such that $\pi(S) = B$. Then Φ is non-void, partially ordered by set inclusion, and every decreasing chain has a lower bound. It follows from Zorn's Lemma that Φ has a smallest element K(0). We show that;

- $\pi|_{K(0)}$ is an open mapping ;
- $\pi|_{K(0)}$ is bijective

This will prove the result for the unique quantum field is then $f = \left(\pi|_{K(0)}\right)^{-1}$, the Axiom of Choice selecting out K(0) as a unique minimal section through the fibre bundle K.

$\pi|_{K(0)}$ is an open mapping ;

Consider then V to be a non-trivial open set in K(0). By definition, $\pi|_{K(0)}$ is surjective, thus; $B \setminus \pi(V) \subset \pi\left(K(0) \setminus V\right)$ . Now $K(0) \setminus V$ is a closed thus compact subset of K, and $\pi|_{K(0)}$ is continuous, thus $\pi\left(K(0) \setminus V\right)$ is compact. It follows that $\left(B \setminus \pi(V)\right)^{-} \subset \pi\left(K(0) \setminus V\right)$. If $\pi(V)$ is nowhere dense, then;

$$B = B \setminus \operatorname{int}\left(\pi(V)\right)^{-} \subset B \setminus \pi(V)^{-} \subset \pi\left(K(0) \setminus V\right)$$, a closed, compact set.

This is a contradiction due to minimality of the set K(0). Thus $\pi(V)$ cannot be a nowhere dense set. From our earlier discussion, this is enough to show that $\pi|_{K(0)}$ is an open mapping.

$\pi|_{K(0)}$ is injective ;

Assume that $\pi|_{K(0)}$ is not injective, then $\exists x_1, x_2; x_1 \neq x_2 \in K(0)$ with $\pi(x_1) = \pi(x_2)$.

Since K(0) is a Hausdorff topological space with a basis of clopen sets, there is a clopen subset V of K(0) containing $x_1$ but not $x_2$. Then;

$\pi(V)$ is also clopen and $\pi^{-1}\left(\pi(V)\right) \setminus V \subset K(0)$, and $\pi(x_2) = \pi(x_1) \in \pi(V)$, thus $x_2 \in \pi^{-1}\left(\pi(V)\right) \setminus V$. It follows that $W = \pi^{-1}\left(\pi(V)\right) \setminus V$ is a non-trivial open set, and W and V are disjoint, with $B \setminus \pi(V) \subset \pi\left(K(0) \setminus W\right)$ and $\pi(V) \subset \pi\left(K(0) \setminus W\right) \Rightarrow \pi\left(K(0) \setminus W\right) = B$

This again contradicts the minimality of K(0). Thus $\pi|_{K(0)}$ is a continuous bijection, and the required lifting is given by $f = \left(\pi|_{K(0)}\right)^{-1}$.

**Pure Gravity States**

We define a *separating* *T*-invariant quantum state $f$ to be a such that given an observable $A$ in the fibre algebra **A(x) ,**

$$f(A^*A) = 0 \Rightarrow A = 0 .$$

We call such states gravity states, motivated by the classical case. If a tensor or the difference between two tensors of the same covariant and contravariant class is equal to zero for all local inertial reference frames, then it is zero for all curvilinear reference frames, by the covariance assumptions of General Relativity.

Since the Hilbert space F(x) on which the fibre algebra **A(x)** acts is separable, by definition, there is a countable dense subset $x(n)$. Defining $f = \sum_1^\infty \alpha_{x(n)}\omega_{x(n)}; \lim_{n\to\infty} \sum_1^n \alpha_{x(j)} = 1$; clearly $f$ is a separating state, and then each element of the weak*-closed convex hull $\Delta = \overline{co}\left\{ f \bullet \alpha_g ; g \in T \right\}$ is also separating, for if;

$$\sum_j \lambda(j) f \circ \alpha(g_j)(A^*A) = 0 \Rightarrow \exists g_j \in T; f \circ \alpha(g_j)(A^*A) = 0 \Rightarrow \alpha(g_j)(A^*A) = 0,$$

since $f$ is separating.

Applying the automorphism $\alpha(g_j^{-1}) \Rightarrow A^*A = 0$. Additionally, if $f_n(A^*A) \to g(A^*A)$ and $f_n(A^*A) = 0$ then $g(A^*A) = 0$. Thus every element of $\Delta$ is separating.

It follows, by applying the Hajian-Kakutani fixed point theorem [6] to the weak*- compact set $\Delta$ that it contains an invariant state. Thus, there is a separating *T*-invariant quantum state.

If $f$ is an extreme point of the separating T-invariant quantum states, let $\pi$ be the GNS representation then clearly $\pi$ is an algebraic isomorphism since the kernel of $f$ is {0}.. Defining;

$$\{U_g ; g \in G, U_g \pi(A)\xi = \pi(\alpha_g(A))\xi\}$$

on the pre-Hilbert space of the GNS representation of $f$; these extend to the 'Segal unitaries' associated with $f$ [4]. The mapping $g \to U_g$ is a unitary representation implementing $\alpha : g \to \alpha_g$. If the mapping $\alpha : g \to \alpha_g$ is weakly measurable in the GNS representation then by our previous results [3, 4] it is norm continuous and the mapping $g \to U_g$ implementing $\alpha : g \to \alpha_g$ is also norm continuous with $U_g \in \pi(\mathbf{A(x)})\ \forall g \in T$

We have, from previous work [7] that the automorphic representation $\alpha : g \to \alpha_g$ of $T$ acts ergodically if and only if $\pi(\boldsymbol{A(x)}) \cap \{U_g ; g \in T\}'$ is trivial, containing only the projections 0 and I and thus consisting of the set of complex multiples of *I*.

We also have;

$$J\left\{\pi(\boldsymbol{A(x)}) \cap \{U_g ; g \in T\}'\right\}J = \pi(\boldsymbol{A(x)})' \cap \{U_g ; g \in T\}' \quad .$$

with $J^2 = 1$. It follows that the representation $\alpha : g \to \alpha_g$ of $T$ acts ergodically if and only if $\pi(\boldsymbol{A(x)})' \cap \{U_g ; g \in T\}'$ is also trivial. But for this case we have also shown that $U_g \in \pi(\mathbf{A(x)})\ \forall g \in T$ and thus if E is a projection in $\pi(\boldsymbol{A(x)})'$ then clearly

$E \in \pi(\boldsymbol{A(x)})' \cap \{U_g ; g \in T\}' = \{\lambda I\}$.

The GNS representation is thus irreducible in the sense of Murray-von Neumann, corresponding to *f* being a pure quantum state.

**The Supersymmetric Massless Graviton**

The Poincare group is a locally compact Lie group with 10 generators, and the translational group is an abelian subgroup generated by the energy-momentum 4-vector $P_\mu$. This has the property that its square $P^2 = P_\mu P^\mu = E^2$ lies in the centre of the Lie algebra. If we consider the energy-momentum vector in normalized units *(c=1)* then $P^2$ has the form $P^2 = m^2 I$, where *m* is the mass-energy of the corresponding particle. In other words, a factorial representation of the Translational group corresponds to a particle with fixed mass *m* and undetermined spin. We can consider two cases;

(a). $P^2 = m^2.I;\ m^2 \neq 0$ corresponding to a multiplet of particles each of the same positive mass but with different spin values;

(b). $P^2 = 0.I$ . This factorial representation corresponds to a massless particle such as a photon or a graviton, with a Supersymmetric massless fermion partner.

**N = 1 Supersymmetry**

For either case we need to add an additional element, normally denoted $Q_\alpha$ , to the Lie algebra, to represent the spread of spin values. In any representation, these are all linear operators, including the identity operator *I*, and thus form an algebra of such operators. Such a factorial representation corresponding to a set of particles, must contain equal numbers of bosons and fermions [8]. With certain assumptions, such a representation where the centre of the algebra is non-trivial can be decomposed into a direct integral of factorial representations, as discussed in [9].

We can develop a locally linear representation of these operators, which is a faithful representation of the Superspace Lie Algebra [8] by adding a pair of Grassmann variables to the algebraic formulation. We can then generate a standard Lie algebra while mixing commutators anticommutators. For example, if $\xi$ and $\bar{\xi}$ have the Grassmann property so that $\xi\bar{\xi} = -\bar{\xi}\xi$ , then; $\left[\xi Q_A, \bar{\xi}\bar{Q}_{\dot{B}}\right] = \left(\xi Q_A \bar{\xi}\bar{Q}_{\dot{B}} - \bar{\xi}\bar{Q}_{\dot{B}}\xi Q_A\right) = \xi\bar{\xi}\left(Q_A\bar{Q}_{\dot{B}} + \bar{Q}_{\dot{B}}Q_A\right) = \xi\bar{\xi}\left\{Q_A, \bar{Q}_{\dot{B}}\right\}$ .

Note that we also require that these Grassmann variables commute with the operators *Q*. A typical element of the corresponding Lie group *G* is then of the form;

$$G(x^\mu, \xi, \bar{\xi}) = \exp i(\xi Q + \bar{\xi}\bar{Q} - x^\mu P_\mu)$$

Here, $x^\mu$ is an event point in locally flat space-time, thus we can think of the Grassmann variables as a vector at the point $x^\mu$ . With this structure, the Superspace is a vector bundle and the locally flat group multiplication structure is of the following form;

$G(x^\mu, \theta, \bar{\theta})G(a^\nu, \xi, \bar{\xi}) = G(x^\mu + a^\nu - i\xi\sigma^\mu\bar{\theta} + i\theta\sigma^\mu\bar{\xi}, \theta + \xi, \bar{\theta} + \bar{\xi})$ . This follows from the Grassmannian properties, since $\theta^2 = \bar{\theta}^2 = 0$, for example.

We can interpret this group product as shifting the locus of the Grassmann vector in space-time from $x^\mu$ to $x^\mu + a^\mu - i\xi\sigma^\mu\bar{\theta} + i\theta\sigma^\mu\bar{\xi}$ together with additive change to the Grassmann vector at this point. If this shift is infinitesimal, then, as in normal Lie group theory, we can consider the tangent plane around the group element $G(a^\mu, \xi, \bar{\xi})$ , giving the following local representation of the Lie group generators on the tangent plane to the Riemannian space-time manifold at the event point $(a^{\nu})$:

$$P_{\mu}(a^{\nu}) = i\frac{\partial}{\partial x^{\mu}}|_{(a^{\nu})} = i\partial_{\mu}|_{(a^{\nu})};$$
$$iQ_{A}(a^{\nu}) = \frac{\partial}{\partial\theta_{A}}|_{(a^{\nu})} - i\sigma^{\mu}\bar{\theta}\frac{\partial}{\partial x^{\mu}}|_{(a^{\nu})}; i\bar{Q}_{\dot{A}}(a^{\nu}) = -\frac{\partial}{\partial\bar{\theta}_{\dot{A}}}|_{(a^{\nu})} + i\theta\sigma^{\mu}\frac{\partial}{\partial x^{\mu}}|_{(a^{\nu})}$$

In this form the generators satisfy all the algebraic relationships of the Supersymmetric extension of the local translation Lie algebra. Thus, this representation is locally an algebraic isomorphism onto the curved Riemannian manifold *M a*nd the piecewise local representations;

$$P_{\mu}(a^{\nu}) = i\frac{\partial}{\partial x^{\mu}}|_{(a^{\nu})} = i\partial_{\mu}|_{(a^{\nu})}$$

are the generators of the local relativistic diffeomorphisms around the event points of *M.* The manifold is assumed smoothly differentiable; we focus on the flat tangent plane and assume Dirac relativistic spinor theory applies. We therefore assume that this operator representation acts on a 4-dimensional Hilbert space *H* of (spinor) wave functions; and we denote by $\psi$ an element of the Hilbert space.

Within these assumptions, from earlier, we can express this state $\psi$ in the form;

$$\psi = \frac{1}{2}(I_{4x4} - \gamma^{5})\psi + \frac{1}{2}(I_{4x4} + \gamma^{5})\psi = \psi_{L} + \psi_{R}.$$

If we define the Dirac adjoint function, $\bar{\psi} = \psi^{+}\gamma^{0}$ where $\psi^{+} = (\psi^{*})^{T}$ then taking complex conjugates of both sides of the massless Dirac equation followed by transposition yields the identity:

$$(\psi^{*})^{T}(-i(\gamma^{\mu *})^{T}(\partial_{\mu} + ieA_{\mu}) = 0$$

Exploiting the fact that $(\gamma^{0})^{2} = I_{4x4}$ leads to the equation:

$$-i\gamma^{\mu T}(\partial_\mu + ieA_\mu)\bar{\psi}^T = 0$$

The matrices $(-\gamma^{\mu T})$ also satisfy the Clifford algebra relations and there is a 4x4 non-singular matrix $C$ such that $C^{-1}\gamma^\mu C = -\gamma^{\mu T}$. We can thus define the charge conjugate spinor

$$\psi^c = C\bar{\psi}^T .$$

In the Weyl representation we take;

$C = i\gamma^0\gamma^2 = i\begin{pmatrix} -\sigma^2 & 0 \\ 0 & \sigma^2 \end{pmatrix}$ where $\sigma^2$ is the second Pauli matrix $\begin{pmatrix} 0 & -i \\ i & 0 \end{pmatrix}$

With $\psi$ the spinor wave function $\begin{pmatrix} \varphi \\ \bar{\chi} \end{pmatrix} = \begin{pmatrix} \varphi_A \\ \bar{\chi}^{\dot{A}} \end{pmatrix}$ we have

$$\psi^c = C\bar{\psi}^T = C(\psi^+\gamma^0)^T = i\gamma^0\gamma^2\gamma^{0T}\psi^* = i\begin{pmatrix} 0 & -\sigma^2 \\ \sigma^2 & 0 \end{pmatrix}\begin{pmatrix} \varphi^* \\ \bar{\chi}^* \end{pmatrix} = i\begin{pmatrix} -\sigma^2\bar{\chi}^* \\ \sigma^2\varphi^* \end{pmatrix}$$

Now note that;

$\varphi^* = (\varphi_A)^* = \bar{\varphi}_{\dot{A}}$ and $\chi = \chi^A = (\bar{\chi}^{\dot{A}})^*$

Thus

$$\psi^c = i\begin{pmatrix} -\sigma^2\bar{\chi}^* \\ \sigma^2\varphi^* \end{pmatrix} = \begin{pmatrix} \begin{pmatrix} 0 & -1 \\ 1 & 0 \end{pmatrix}\bar{\chi}^* \\ \begin{pmatrix} 0 & 1 \\ -1 & 0 \end{pmatrix}\varphi^* \end{pmatrix} = \begin{pmatrix} \varepsilon_{AB}(\bar{\chi}^{\dot{A}})^* \\ \varepsilon^{\dot{A}\dot{B}}(\varphi_A)^* \end{pmatrix} = \begin{pmatrix} \varepsilon_{AB}\chi^A \\ \varepsilon^{\dot{A}\dot{B}}\bar{\varphi}_{\dot{A}} \end{pmatrix} = \begin{pmatrix} \chi_B \\ \bar{\varphi}^{\dot{B}} \end{pmatrix} \in H$$

where the $\varepsilon$ matrices are the spinor metric 2x2 matrices.

**The Mass Zero Case as a Graded Lie Algebra**

We start with the properties of a $Z_2$-graded Lie algebra $L = L_0 \oplus L_1$ where $L_0$ is the Lie algebra spanned by the generators $P_\mu$ $(\mu = 0,1,2,3)$ and
$L_1$ is spanned by the spinor charge generators $\mathrm{Q}_\alpha$ $(\alpha = 0,1,2,3)$.

From the definition of a $Z_2$-graded Lie algebra $L$;

$P_\mu \in L_0, Q_\alpha \in L_1$ with gradings 0 and 1
$(P_\mu, Q_\alpha) \to P_\mu \circ Q_\alpha = P_\mu Q_\alpha - (-1)^{0x1} Q_\alpha P_\mu = P_\mu Q_\alpha - Q_\alpha P_\mu = [P_\mu, Q_\alpha] = 0$

Similarly,

We also have $\mathrm{Q}_\alpha \circ Q_\beta = \mathrm{Q}_\alpha Q_\beta - (-1)^{1x1} Q_\beta \mathrm{Q}_\alpha = \mathrm{Q}_\alpha Q_\beta + Q_\beta \mathrm{Q}_\alpha = \{\mathrm{Q}_\alpha, Q_\beta\}$, the anticommutator.

Since $Q_\alpha$ is a non-Hermitian operator (by construction), we can also consider the complex conjugate Dirac 4-spinor $Q_\beta \in \dot{F}$ . To resolve differences in the literature we assume $\bar{Q}_\beta = \bar{Q}_{\dot{\beta}}$ .

To constrain the number of options it is convenient at this stage to assume;

$Q = Q^c = C\bar{Q}^T$
Thus $(CQ)^T = (C^2 \bar{Q}^T)^T = -\bar{Q}$ since $C^2 = -1$.
Hence $Q^T C = \bar{Q}$ since $C^T = -C$

Following now the logic of [8] in general, we consider from the $Z_2$ grading,

$\{Q_\alpha, Q_\beta\} = Q_\alpha Q_\beta + Q_\beta Q_\alpha = a(\gamma^\mu C)_{\alpha\beta} P_\mu$
Multiplying from the right by $C$; $Q_\alpha Q_\beta C_{\beta\delta} + Q_\beta C_{\beta\delta} Q_\alpha = a(\gamma^\mu)_{\alpha\gamma} C_{\gamma\beta} C_{\beta\delta} P_\mu = -a(\gamma^\mu)_{\alpha\delta} P_\mu$

Hence $Q_\alpha (Q^T C)_\delta + (Q^T C)_\delta Q_\alpha = -a(\gamma^\mu)_{\alpha\delta} P_\mu$
Thus $\{Q_\alpha, \bar{Q}_\beta\} = -a(\gamma^\mu)_{\alpha\delta} P_\mu$

We assume the operators $Q, \bar{Q}$ are Marjorana 4-spinors; then in 2-spinor notation we can simply replace the Dirac $\gamma$ matrices with their equivalent Pauli matrices yielding the following relationship;

$$\{Q_A, \bar{Q}_{\dot{B}}\} = -a(\sigma^\mu)_{A\dot{B}} P_\mu \qquad (2)$$

Similarly, we can show that, for 4-spinors, $\{\bar{Q}_\alpha, \bar{Q}_\beta\} = -a(C^{-1}\gamma^\mu)_{\alpha\beta} P_\mu$

Hence in 2-spinor notation, $\{\bar{Q}_A, \bar{Q}_{\dot{B}}\} = -a(C^{-1}\sigma^\mu)_{A\dot{B}} P_\mu$

Since all the latter equations are relativistically invariant, we can transform them to the rest frame where $P_\mu = (E,0,0,0) = (m,0,0,0)$ with the speed of light normalised at $c = 1$.

With these values of $P_\mu$ in equation (2) above we have

$\{Q_A, \bar{Q}_{\dot{B}}\} = -a\sigma^0_{A\dot{B}} P_0$

Hence $\{Q_A, \bar{Q}_{\dot{B}}\}\sigma^0_{\dot{B}A} = -a\sigma^0_{A\dot{B}}\sigma^0_{\dot{B}A} P_0 = -aP_0$

Taking $A = 1,\ \dot{B} = \dot{1}$ we have $\{Q_1, \bar{Q}_{\dot{1}}\}\sigma^0_{\dot{1}1} = -aP_0$

Similarly for $A = 2,\ \dot{B} = \dot{2}$ we have $\{Q_2, \bar{Q}_{\dot{2}}\}\sigma^0_{\dot{2}2} = -aP_0$

Since $\sigma^0_{\dot{1}1} = \sigma^0_{\dot{2}2} = 1$ we have $\{Q_1, \bar{Q}_{\dot{1}}\} + \{Q_2, \bar{Q}_{\dot{2}}\} = -2aP_0$

For consistency with the current literature, we assume the constant $a$ = -1.

Thus, we have the quantum operator equality;

$$\{Q_1, \bar{Q}_{\dot{1}}\} + \{Q_2, \bar{Q}_{\dot{2}}\} = 2P_0$$

The left-hand side of this expression is a positive definite quantum operator thus for $\psi$ an element of the Hilbert space;

If $\psi$ is the vacuum state then $<\psi, P_0\psi> = 0$ is equivalent to $2 < \psi, Q_1\bar{Q}_{\dot{1}}\psi > +2 < \psi, Q_2\bar{Q}_{\dot{2}}\psi > = 0$

Since, for 2-spinors, $\varphi^*_A = \bar{\varphi}_{\dot{A}}$ we can rewrite this as: $2 < \psi, Q_1 Q_1^* \psi > +2 < \psi, Q_2 Q_2^* \psi > = 0$

Thus $2\,\|\,Q_1\psi\,\|^2 + 2\,\|\,Q_2\psi\,\|^2 = 0,$ from which we deduce that $Q_1\,|\,\psi> = Q_2\,|\,\psi> = 0$

and also $Q_{\dot{1}}\,|\,\psi> = Q_{\dot{2}}\,|\,\psi> = 0 \Rightarrow \omega_\psi(Q_1) = \omega_\psi(Q_2) = \omega_\psi(Q_{\dot{1}}) = \omega_\psi(Q_{\dot{2}}) = \omega_\psi(P_0) = 0.$

**Factorial Representations of the $Z_2$ graded Lie algebra**

The extension of the space $L_0$ to the space $L = L_0 \oplus L_1$ maintains $P^2$ as an element of the centre;

$$\left[P^2, Q_A\right] = 0 = \left[P^2, \bar{Q}^{\dot{A}}\right]$$

In a factorial representation $\Pi$ of the $Z_2$ graded algebra, it follows that $\Pi(P^2) = m^2 I$, fixing the mass $m$ of all particles in this representation. However, the spins of the particles in this representation are not fixed at a common value.

In this factorial 2-spinor representation, we have, from earlier, the algebraic identity ;

$$\{Q_A, \bar{Q}_{\dot{B}}\} = \sigma^\mu_{A\dot{B}} P_\mu .$$

With $P^2 = m^2 I$ in this representation, and setting $P_\mu = (m,0,0,0)^T$ in the rest frame, we have the following set of identities from equation (2) taking $A = 1,2$ and $\dot{B} = \dot{1},\dot{2}$;

$$\{Q_1, \bar{Q}_{\dot{1}}\} = \sigma^{\mu}_{1\dot{1}} P_\mu = \sigma^{0}_{1\dot{1}} m = m$$
$$\{Q_1, \bar{Q}_{\dot{2}}\} = \sigma^{\mu}_{1\dot{2}} P_\mu = \sigma^{0}_{1\dot{2}} m = 0$$
$$\{Q_A, Q_B\} = \{\bar{Q}_{\dot{A}}, \bar{Q}_{\dot{B}}\} = 0$$

From these properties we see that the 2-spinors form at least a Clifford algebra in this factorial representation, and we see also that $Q_A^2 = 0$ and this is in fact a Grassmann algebra.

If $|\, p, \lambda >$ is an eigenstate in the Hilbert space *H*, then $P_\mu \,|\, p, \lambda >= p_\mu \,|\, p, \lambda >$.

The corresponding (pure) vector state is invariant under the translational group since we have

$$0 = P_\mu Q_A \,|\, p,\lambda >= Q_A P_\mu \,|\, p,\lambda >= Q_A p_\mu \,|\, p,\lambda >= p_\mu Q_A \,|\, p,\lambda >$$

We can thus always choose a minimum energy pure vector state $\omega_{|p,\lambda>}$ which is translation invariant and with $\omega|_{|p,\lambda>}(Q_A) =< p,\lambda \,|\, Q_A \,|\, p,\lambda >= 0$. It is specified by its mass-energy *p* and its spin value $\lambda$. Thus $\omega_{|p,\lambda>}$ is a translation invariant ergodic pure state.

If we now, exploiting local special relativistic covariance, choose an inertial reference frame in which the Wigner little group contains the spin generating 2x2 rotation matrices in the x-y plane, we have;

$P_\mu = (E,0,0,E)$ therefore;

$$\{Q_A, \bar{Q}_{\dot{B}}\} \,|\, p,\lambda >= \sigma^{\mu}_{A\dot{B}} P_\mu \,|\, p,\lambda >= \sigma^{\mu}_{A\dot{B}} p_\mu \,|\, p,\lambda >$$

Now $\sigma^{\mu}_{A\dot{B}} p_\mu = \sigma^0 p_0 - \sigma^1 p_1 - \sigma^2 p_2 - \sigma^3 p_3 = E(\sigma^0 - \sigma^3)$

$$= E\left( \begin{pmatrix} 1 & 0 \\ 0 & 1 \end{pmatrix} - \begin{pmatrix} 1 & 0 \\ 0 & -1 \end{pmatrix} \right) = \begin{pmatrix} 0 & 0 \\ 0 & 2E \end{pmatrix}$$

Hence applying this to our translation invariant vacuum state, $\omega_{|p,2>}$ we have;

$|< p,2|Q_1\bar{Q}_{\dot{1}}|p,2>=-< p,2|\bar{Q}_{\dot{1}}Q_1|p,2>=0$

Similarly, considering the $2\dot{1}$ element of the matrix, we have

$|< p,2|Q_2\bar{Q}_{\dot{1}}|p,2>=-< p,2|\bar{Q}_{\dot{1}}Q_2|p,2>=0$

Thus $(aQ_1+bQ_2)\bar{Q}_{\dot{1}}|p,2>=0$ for any scalars $a$ and $b$.

Since $Q_1$ and $Q_2$ span the subspace $L_1$, it follows that $\bar{Q}_{\dot{1}}|p,2>=0$.

We conclude that the translation invariant pure Boson state $\omega_{|p,2>}$ is the local Clifford vacuum, and a spin 2 pure state corresponding to the graviton. It is an extreme point of the closed convex hull of the state space and is thus an extreme point of the translation invariant states: it is an ergodic gravity state. Its Supersymmetric fermion partner is the Gravitino with spin $\frac{3}{2}$.

## Quantum States Invariant Under the Action of Compact Lie Groups

The weak topology $\sigma(A(x)_*, A(x))$ can be defined on the predual $A(x)_*$ as the coarsest topology for which elements of the predual are continuous [10]. It is defined by a set of semi-norms $p = |f|$ for $f$ a density matrix linear functional which as a set are separating for $A(x)_*$. Making minimal assumptions we let $\alpha : g \to \alpha_g$ be a weakly measurable representation of the compact Lie group $G$ as automorphisms of **A(x)**. By this we mean that the induced mapping[2] $\nu : g \to f \circ \alpha_g^{-1} : G \to \mathbf{A(x)}_*$ is measurable for Haar measure on $G$ and the $\sigma(A(x)_*, A(x))$ topology on $A(x)_*$ Since every positive element of $\mathbf{A(x)}_*$ is a countable sum of vector states this is equivalent to the definition that $\nu : g \to \omega_x \circ \alpha_g : G \to \mathbf{A(x)}_*$ is measurable for all $x$ in the fibre Hilbert space *F(x).*

Given that the induced mapping $\nu : g \to f \circ \alpha_g : G \to \mathbf{A(x)}_*$ is measurable in the sense now defined above we have from [11];

$$\|\nu(g)\circ f(A) - \nu(h)\circ f(A)\| \le \|\nu(g)\circ f - \nu(h)\circ f\|\,\|A\| \to 0 \text{ as } g \to h$$

---

[2] We ignore the inverse symbol in this definition to ease notational clutter.

This demonstrates the following result, which allows the extension of continuous gauge automorphic representations of compact Lie groups to their cross- product such as the Standard Model gauge group $SU(3)\times SU(2)\times S(1)$;

*For the induced representation* $v: g \to f \circ \alpha_g : G \to$ **A(x)**$_*$ *on the predual of* ***A(x),*** *weak measurability is equivalent to weak continuity.*

We have shown, as for local diffeomorphism-invariant quantum states [3, 7], that quantum states invariant under the action now of compact Lie groups are common in the sense that the weakly closed convex hull of every normal state contains such a state. We are now dealing with groups such as *SU(n)* which are both compact and non-abelian thus different techniques are required. To achieve this result, we developed a new idea based on group stabilizer theory which we called Wigner sets [12]. These are complementary to little groups.

**Wigner Sets and the Finite Intersection Property**

Given a density matrix quantum state *f*, and a weakly measurable representation $g \to \alpha_g$ of a compact Lie group *G* as gauge automorphisms of the fibre algebra **A(x)**; define the closed convex hull; $X(f) = \overline{co}\left\{ f \circ \alpha_g ; g \in G \right\} \subset$ **A(x)**$_*$ with closure in the $\sigma\left(A(x)_*, A(x)\right)$-topology. Define the group of isometric and $\sigma\left(A(x)_*, A(x)\right)$-continuous transformations mapping $X(f)$ $\to X(f)$ by $v(G) = \left\{ v(g) : x \to x \circ \alpha_g ; g \in G, x \in X(f) \right\}$.

Mathematically, we note that since *G* is compact and $f \circ \alpha : G \to$ **A(x)**$_*$ is weakly measurable and thus weakly continuous; this implies that $f \circ \alpha(G)$ is $\sigma\left(A(x)_*, A(x)\right)$-compact. The Krein - Smulian theorem [13] , then shows that *X( f )* is also a $\sigma\left(A(x)_*, A(x)\right)$- compact set. Thus *X( f )* is a non-void $\sigma\left(A(x)_*, A(x)\right)$-compact convex subset of the locally convex Hausdorff linear topological space of ultraweakly continuous linear functionals acting on the fibre algebra **A(x)** .

The group of mappings $v(G) = \left\{ v(g) : x \to x \circ \alpha_g ; g \in G, x \in X(f) \right\}$ is a non-contracting (semi)-group of weakly continuous affine maps of *X( f )* onto itself. We can therefore, apply the Ryll-Nardzewski fixed point theorem [14] to establish the existence of an invariant normal state contained in *X( f ).*

The physical implications highlight the role of what we have termed Wigner sets.

Given $g \in G,$ define the Wigner set of the mapping $\nu(g): X\left(f\right) \to X\left(f\right)$ as the stabiliser set; $\mathcal{F}\left(\nu(g)\right) = \left\{x \in X(f); \nu(g) \circ x = x\right\}.$

More generally, given a finite subset $\left\{g(j) \in G; j =, 2, ...n\right\}$ and corresponding mappings $\left\{\nu\left(g\left(j\right)\right); j =, 2, ...n\right\}$ , we can construct the affine mapping $\frac{1}{n}\left(\sum_j \nu\left(g\left(j\right)\right)\right): X(f) \to X(f).$

We then define $\mathcal{F}\left\{\nu\left(g\left(j\right)\right); j =, 2, ...n\right\} \triangleq \mathcal{F}\left(\frac{1}{n}\left(\sum_j \nu\left(g\left(j\right)\right)\right)\right)$

We proved [12] that the following relationship between Wigner sets applies;

$$\bigcap_j \mathcal{F}\left(\nu\left(g\left(j\right)\right)\right) = \mathcal{F}\left(\frac{1}{n}\left(\sum_j \nu\left(g\left(j\right)\right)\right)\right)$$

**Invariant Normal States**

It is now easy to see why, physically, by exploiting the properties of Wigner sets, *X( f )* contains a fixed point for the group of isometric and $\sigma\left(A(x)_*, A(x)\right)$-continuous transformations $\nu\left(G\right) = \left\{\nu\left(g\right)(x) = x \circ \alpha_g; g \in G, x \in X(f)\right\}.$

We have that $\left(\frac{1}{n}\left(\sum_j \nu\left(g\left(j\right)\right)\right)\right)$ is a $\sigma\left(A(x)_*, A(x)\right)$- continuous affine mapping on the compact convex set *X( f )* It thus has a fixed point $x$ (applying again Schauder's fixed point theorem). Then $x \in \mathcal{F}\left(\frac{1}{n}\left(\sum_j \nu\left(g\left(j\right)\right)\right)\right).$

The expression;

$$\bigcap_{j}\mathcal{F}\left(\nu\left(g\left(j\right)\right)\right)=\mathcal{F}\left(\frac{1}{n}\left(\sum_{j}\nu\left(g\left(j\right)\right)\right)\right)$$

shows that the Wigner sets $\mathcal{F}\left(\nu\left(g\right)\right)$ have the finite intersection property since $\bigcap_{j}\mathcal{F}\left(\nu\left(g\left(j\right)\right)\right)$ is non-void . Clearly each Wigner set is a ($\sigma\left(A(x)_{*}, A(x)\right)$) closed subset of the compact set $X(f)$ thus $\bigcap_{g}\mathcal{F}\left(\nu(g)\right)\neq\varnothing$. If $h\in\bigcap_{g}\mathcal{F}\left(\nu(g)\right)\Rightarrow h=\nu_{g}\circ h=h\circ\alpha_{g}\forall g\in G$. Thus, *h* is the required invariant quantum state.

The proof shows that quantum states invariant under the action of compact Lie groups are common in the sense that the weakly closed convex hull of every normal state contains such an invariant state.

**Discussion**

We have developed a set of mathematically consistent non perturbative methods applicable to Loop Quantum Gravity, addressing a number of issues raised by Ashtekar [15]. These methods have been applied to show that quantum states invariant under either an external group of local diffeomorphisms of space-time or quantum states invariant under the internal action of a compact Lie group are 'common', in the sense that the weakly closed convex hull of every relevant quantum state contains such an invariant state. These form the building blocks of invariant fields and Lagrangians. A form of $N$ = 1 Supersymmetry and non-commutative space – time naturally emerges, which predicts a spin- 2 massless graviton and its companion gravitino. Separately, in [16] these methods are used to develop a quantum ergodic theory.